\documentclass[a4paper,twocolumn,11pt,accepted=2021-07-22]{quantumarticle}
\pdfoutput=1
\usepackage[utf8]{inputenc}
\usepackage[english]{babel}
\usepackage[T1]{fontenc}
\usepackage{amsmath}
\usepackage{hyperref}
\usepackage{amsthm}

\usepackage[numbers,sort&compress]{natbib}
\usepackage{graphicx}
\usepackage{dcolumn}
\usepackage{amssymb}
\usepackage{mathtools}
\usepackage{float}
\usepackage{bm}

\theoremstyle{plain}
\newtheorem{thm}{Theorem}

\newtheorem{defin}{Definition}
\newtheorem{lem}{Lemma}
\newtheorem{cor}{Corollary}
\newtheorem{prop}{Proposition}

\newcommand\ind{\protect\mathpalette{\protect\independenT}{\perp}}
\def\independenT#1#2{\mathrel{\rlap{$#1#2$}\mkern2mu{#1#2}}}

\newcommand*{\QED}{\hfill\ensuremath{\square}}

\newcommand{\mc}[1]{\mathcal{#1}}

\newcommand{\bq}{\begin{equation}}
\newcommand{\eq}{\end{equation}}
\newcommand{\ba}{\begin{align}}

\newcommand{\Ae}{\bm{A}_{\setminus i}}
\newcommand{\Xe}{\bm{X}_{\setminus i}}

\newcommand{\Ce}{\bm{C}_{\setminus i}}

\newcommand{\Ze}{\bm{Z}_{\setminus i}}
\newcommand{\Zj}{\bm{Z}_{\setminus j}}

\newcommand{\Le}{\bm{\Lambda}}
\newcommand{\Om}{\bm{\Omega}}
\newcommand{\Xib}{\bm{\Xi}}
\newcommand{\Agm}{\bm{A}_{\setminus \gamma}}
\newcommand{\Xgm}{\bm{X}_{\setminus \gamma}}
\newcommand{\Ag}{\bm{A}_{\gamma}}
\newcommand{\Xg}{\bm{X}_{\gamma}}

\newcommand{\Pa}{\mathrm{Pa}}

\newcommand{\Nd}{\mathrm{Nd}}

\newcommand{\expect}[1]{\left\langle#1\right\rangle} 

\usepackage{color}

\definecolor{red}{rgb}{0.9,0,0}
\definecolor{blue}{rgb}{0,0,0.8}
\definecolor{ngreen}{rgb}{0.3,0.6,0.2}
\definecolor{pink}{rgb}{1.0,0.3,0.7}

\begin{document}
	
\title{Classical causal models cannot faithfully explain Bell nonlocality or Kochen-Specker contextuality in arbitrary scenarios}

\author{J. C. Pearl}
\email{jason.pearl@griffithuni.edu.au}
\affiliation{Centre for Quantum Dynamics, Griffith University, Gold Coast, QLD 4222, Australia}

\author{E. G. Cavalcanti}
\email{e.cavalcanti@griffith.edu.au}
\affiliation{Centre for Quantum Dynamics, Griffith University, Gold Coast, QLD 4222, Australia}

\begin{abstract}
	In a recent work, it was shown by one of us (EGC) that Bell-Kochen-Specker inequality violations in phenomena satisfying the no-disturbance condition (a generalisation of the no-signalling condition) cannot in general be explained with a faithful classical causal model---that is, a classical causal model that satisfies the assumption of \emph{no fine-tuning}. The proof of that claim however was restricted to Bell scenarios involving 2 parties or Kochen-Specker-contextuality scenarios involving 2 measurements per context. Here we show that the result holds in the general case of arbitrary numbers of parties or measurements per context; it is not an artefact of the simplest scenarios. This result unifies, in full generality, Bell nonlocality and Kochen-Specker contextuality as violations of a fundamental principle of classical causality. We identify, however, an implicit assumption in the former proof, making it explicit here: that certain operational symmetries of the phenomenon are reflected in the model, rather than requiring fine-tuned choices of model parameters. This clarifies a subtle but important distinction between Bell nonlocality and Kochen-Specker contextuality.
\end{abstract}


\maketitle

\section{Introduction}\label{sec:intro}
Bell nonlocality \cite{Bell1964} and Kochen-Specker (KS) contextuality \cite{Kochen1967} are classically forbidden correlations characteristic of quantum phenomena. Bell nonlocality can be understood as the impossibility to explain certain quantum correlations between space-like separated systems within a classical theory of causality, assuming relativistic causal structure \cite{Wiseman2017}. KS-contextuality, on the other hand, can be understood, within the framework of ontological models~\cite{Spekkens2005}, as the incompatibility between the predictions of quantum theory with the joint assumption of measurement noncontextuality---the assumption that the outcome statistics of a phenomenon should not depend on the measurement context---and outcome determinism.

The fundamentally quantum nature of contextual and nonlocal correlations lies at the heart of many quantum protocols. Bell nonlocality is a key resource for quantum communication, with applications such as reducing communication complexity \cite{Brukner2004} and secure communication \cite{Ekert1992}. Since classical simulation of Bell correlations is possible (between time-like separated system) via the addition of communication channels between the parties in the Bell test~\cite{Brassard1999,Toner2003}, quantum over classical advantages provided by Bell nonlocality can be understood as quantum protocols having access to correlations that can only be simulated classically with the aid of extra resources. KS-contextuality, on the other hand, has been identified as a key resource fuelling quantum over classical advantages in quantum computation \cite{Anders2009, Raussendorf2013, Howard2014, Mansfield2018}.

A modern approach is to encode correlations for a set of observed variables in the framework of causal models, where a causal structure is represented as a directed acyclic graph (DAG)~\cite{Pearl2000, Wood2015}. Recently, a framework was introduced to unify KS-contextuality and Bell nonlocality as violations of a fundamental principle of causal models: the principle of \emph{no-fine-tuning}, or \emph{faithfulness}~\cite{Cavalcanti2018}. In the framework of causal models, fine-tuning occurs when specific choices of parameters of the model (such as distributions over latent variables) hide from operational accessibility some causal connections available in the model. In~\cite{Wood2015} it was shown that representing certain Bell-inequality violations by classical causal models requires fine-tuning, and this result was extended to the case of KS contextuality in~\cite{Cavalcanti2018}. 
Considering a classical causal model to be (essentially) a classical simulation of a quantum phenomenon, this provides a novel approach to understanding the quantum over classical advantage provided by Bell-KS correlations: fine-tuning can be considered an \emph{unavoidable resource waste} in any classical simulation, relative to the quantum realisation of the same correlations. This causal perspective also reinforces the program of revising the assumptions underlying the classical causal models framework, such as Reichenbach's principle of common cause \cite{Reichenbach1956, Cavalcanti2014}---towards a general framework of quantum causal models \cite{Cavalcanti2014, Henson2014, Pienaar2015, Chaves2015, Costa2016, Allen2016, Barrett2019}.

The proofs that classical causal models for Bell-KS correlations require fine-tuning, however, are so far restricted to bipartite Bell scenarios~\cite{Wood2015} or KS scenarios with two measurements per context~\cite{Cavalcanti2018}. As quantum protocols can make use of large numbers of parties or measurements per context, a general proof is needed for this approach to have practical merit. Here we generalise the framework of~\cite{Cavalcanti2018} to arbitrary numbers of parties or measurements per context, demonstrating in full generality the need for fine-tuning in classical causal models for Bell-KS inequality violations.

In the present work, we also correct a subtle but important issue in the definition of no fine-tuning used in \cite{Cavalcanti2018}. That definition did not account for the possibility that the same measurement could have different statistics depending on which random variable in a causal model it is associated with, which would represent a form of contextuality not ruled out by the notion of no fine-tuning used in that work. Here we find that an updated definition can correct this issue, by including the requirement that operational symmetries of the phenomenon must be reflected in the model -- a requirement analogous to the notion of ``operational no fine-tuning'' recently introduced by Catani and Leifer~\cite{Catani2020}.

This paper is organised as follows. In Section~\ref{sec:cms} we give a brief review of the formalism of classical causal models, and Section~\ref{sec:framework} then sets up a framework for describing Bell-KS contextuality scenarios within that formalism. The main result of this work is then given in Section~\ref{sec:main}. In Section~\ref{sec:example}, an example is given to demonstrate how to translate a well-known Kochen-Specker scenario (the Peres-Mermin square) into the causal framework. A technical proof of the main result is provided in Section~\ref{sec:proofs}. We conclude by discussing some important implications of this work, as well as providing some possible avenues for future research in Section~\ref{sec:conclusion}.

\section{Causal models}\label{sec:cms}
Causal models have been developed as a tool for connecting causal inferences and probabilistic observations, with a wide range of applications, from statistics to epidemiology, economics and computer science~\cite{Pearl2000}. In this framework, a causal structure is represented by a graph $\mc{G}$ containing a set of observable variables of interest, as well as additional latent, or hidden, variables. Variables are represented as nodes, with causal links represented by directed edges (arrows). For a pair of variables $\{A,B\}$, $A$ is considered to be the direct cause of B should the graph $\mc{G}$ contain a directed edge from $A$ to $B$. Topologically ordered directed graphs (i.e.~those that exclude the possibility of paradoxical causal loops) are known as directed acyclic graphs (DAGs).

Standard terminology will be used to refer to relationships between variables. If there is is a directed path from $A$ to $B$, then $A$ is said to be an ancestor of $B$, and $B$ is a descendent of $A$. If $A$ has a directed edge to $B$ (i.e.~$A$ is a direct cause of $B$), then $A$ is said to be the parent of $B$. The set of all parents (all direct causes) for $B$ is denoted by $\Pa(B)$; the set of all non-descendents of $B$ is denoted by $\Nd(B)$. The \emph{Causal Markov Condition} is the assumption that a variable $X$ is conditionally independent of its non-descendents, given its parents. This conditional independence (C.I.) is denoted as $(X\ind \Nd(X)\mid \Pa(X))$, meaning that $P(X\mid \Nd(X),\Pa(X))=P(X\mid \Pa(X))$. For a DAG $\mc{G}$ containing variables $\{X_1,\dots,X_n\}$, the Causal Markov Condition implies that any probability distribution \emph{compatible} with $\mc{G}$ factorises as
\begin{equation}\label{GrepresentsP}
P(X_1,\dots,X_n) = \prod_j P(X_j\mid \Pa(X_j)).
\end{equation}

A procedure called \emph{d-separation} (directional separation) can be used to obtain C.I.~relations from a graph \cite{Pearl2000}. Here and henceforth we use a bold symbol to refer to (variables associated with) a set of nodes in a graph. A path $p$ connecting a set of nodes $\bm{X}$ with a set of nodes $\bm{Y}$ is blocked (d-separated) by a set of nodes $\bm{Z}$ if and only if
\begin{enumerate}
	\item $p$ contains a chain $ A\rightarrow B\rightarrow C $ or a fork $ A\leftarrow B\rightarrow C $ such that the middle node $B$ is in $\bm{Z}$, or
	\item $p$ contains an inverted fork (collider) $ A\rightarrow B\leftarrow C $ such that the middle node $B$ is not in $\bm{Z}$ and such that no descendant of $B$ is in $\bm{Z}$. 
\end{enumerate}
A set $\bm{Z}$ is said to d-separate $\bm{X}$ from $\bm{Y}$ (denoted $(\bm{X}\ind \bm{Y}\mid \bm{Z})_d$ if and only if $\bm{Z}$ blocks every path from a node in $\bm{X}$ to a node in $\bm{Y}$. 

The d-separation condition is a \emph{sound} and \emph{complete} criteria for conditional independence. Sound: if the d-separation condition $(\bm{X}\ind \bm{Y}\mid \bm{Z})_d$ is satisfied by a graph $\mc{G}$, then all probability distributions compatible with $\mc{G}$ satisfy the C.I.~relation $(\bm{X}\ind \bm{Y}\mid \bm{Z})$; complete: if all probability distributions compatible with $\mc{G}$ satisfy $(\bm{X}\ind \bm{Y}\mid \bm{Z})$, then $\mc{G}$ satisfies the d-separation condition $(\bm{X}\ind \bm{Y}\mid \bm{Z})_d$. Note that d-separation refers to a relation between $\bm{X}$, $\bm{Y}$ and $\bm{Z}$ relative to a graph, and can also be applied to a subgraph $\mc{S}$ of a graph $\mc{G}$. A d-separation condition obeyed by a subgraph $\mc{S}$ is not necessarily obeyed by $\mc{G}$ however. 

Conditional independence relations satisfy certain properties called \emph{semi-graphoid axioms}~\cite{Pearl2000}: \\
\textit{Symmetry,}
\begin{equation}\label{symmetry}
(\bm{X}\ind \bm{Y}\mid \bm{Z})\bm{\Leftrightarrow} (\bm{Y}\ind \bm{X}\mid \bm{Z})\,,
\end{equation}
\textit{Decomposition,}
\begin{equation}\label{decomposition}
(\bm{X}\ind \bm{YW}\mid \bm{Z})\bm{\Rightarrow} (\bm{X}\ind \bm{Y}\mid \bm{Z})\,,
\end{equation}
\textit{Weak union,}
\begin{equation}\label{weakunion}
(\bm{X}\ind \bm{YW}\mid \bm{Z})\bm{\Rightarrow} (\bm{X}\ind \bm{Y}\mid \bm{ZW})\,,
\end{equation}
\textit{Contraction,}
\begin{multline}\label{contraction}
(\bm{X}\ind \bm{Y}\mid \bm{Z})\textbf{ \& } (\bm{X}\ind \bm{W}\mid \bm{ZY})\\
\bm{\Rightarrow} (\bm{X}\ind \bm{YW}\mid \bm{Z})\,.
\end{multline}

\section{Causal framework for Bell-Kochen-Specker contextuality \& nonlocality}\label{sec:framework} The framework used here generalises that of \cite{Cavalcanti2018}, where traditional ontological models for Bell-nonlocality and contextuality were translated into the language of causal models. Some of the terminology follows that of~\cite{Abramsky2011}.

A \emph{measurement scenario}, or \emph{contextuality scenario}, or \emph{compatibility scenario} is specified by a set of $k$ measurements $\mc{M}=\{m_1,\dots,m_k\}$, a set $\mc{O}$ of possible outcomes for each measurement, and a \emph{compatibility structure} $\mc{C}$, defined to contain all subsets of jointly measurable members of $\mc{M}$: a subset $c\subseteq \mc{M}$ is said to be \emph{jointly measurable}, \emph{compatible}, or to represent a \emph{measurement context} iff $c\in \mc{C}$. A special class of contextuality scenarios are $n$-partite \emph{Bell-nonlocality scenarios}, where $\mc{M}$ can be decomposed into $n$ disjoint subsets $\mc{M}=\{\mc{M}_1,\dots,\mc{M}_n\}$ such that each context $c\in \mc{C}$ contains exactly one element from each subset. We define a \emph{Kochen-Specker (KS) scenario} as any contextuality scenario that is \emph{not} a Bell scenario. We will also refer to an arbitrary (Bell \emph{or} KS) contextuality scenario as a Bell-KS scenario.

Here we consider a general class of measurement scenarios, with no restriction on the number of measurements per context. For simplicity, however, and without loss of generality, we augment all contexts, where needed, with trivial measurements (that always give the same outcome), so that all contexts contain exactly the same number of measurements $n=\mathrm{max}_{c\in \mc{C}}|c|$. Similarly, there is no loss in generality by assigning the same outcome set $\mc{O}$ to every measurement, as $\mc{O}$ can be made large enough to include all possible outcomes of all $m_i\in\mc{M}$. 

Given a measurement scenario, in each \emph{test}, that is, in each run of the experiment, a set of $n$ compatible measurements is chosen to be performed, via a set of random variables $\bm{X}=\{X_1,X_2,...,X_n\}$. That is, in each run the random variables take values so as to form a measurement context, e.g.~$\{X_1=m_1,...,X_n=m_n\}\in \mc{C}$. The respective outcomes are recorded by the set of random variables $\bm{A}=\{A_1,A_2,...,A_n\}$. Measurement-outcome pairs are represented as ordered pairs $(X_i,A_i)$ for all $i\in \mc{I}=\{1,2,...,n\}$. For convenience, we denote an index subset by $\gamma\subseteq \mc{I}$ such that $\Ag\subseteq \bm{A}$ and $\Xg\subseteq \bm{X}$. We then introduce the shortcut notations $\Agm=\bm{A}\setminus \Ag$ and $\Xgm=\bm{X}\setminus \Xg$ for the complementary subsets of variables. We define a test so that for Bell scenarios, each variable $X_i$ (corresponding to the $i^{th}$ party in the test) is always chosen from its corresponding subset $\mc{M}_i$.

Some remarks about the identification of measurements are in order. In Bell scenarios, each measurement choice can be thought of as a setting on a ``black box'' in possession of one of $n$ agents. In KS scenarios, in which the measurement set $\mc{M}$ cannot be factorised that way, some further justification is needed to identify the ``same'' measurement in different contexts. This could be done, following~\cite{Spekkens2005}, via operational equivalence classes. Depending on the implementation, each random variable $X_j$ could then be thought of either as an experimental ``slot'' in a process, or simply as an arbitrary label. For example, $X_j$ could refer to the $j^{th}$ measurement in a temporal sequence. Alternatively, it could refer to the $j^{th}$ measurement to be chosen in an arbitrary way, specifying a joint measurement only after all $n$ measurements are chosen. In either of these cases, the same measurement $m$ can be associated with different random variables $X_j$ in different tests.

A \emph{phenomenon} is specified by a probability distribution $\mc{P}(\bm{AX})$ for all allowed values of the observable variables. Note that the formalism so far is independent of any causal structure. We now define a (classical) causal model for a phenomenon.
\begin{defin}[Classical causal model]\label{defin:CM}
A \emph{classical causal model} $\Gamma$ for a phenomenon $\mc{P}$ consists of a (possibly empty) set of latent variables $\Xib$, a DAG $\mc{G}$ with nodes $\{\bm{A,X},\Xib\}$, and a probability distribution $P(\bm{AX}\Xib)$ compatible with $\mc{G}$, such that $\mc{P}(\bm{AX})=\sum_{\Xib} P(\bm{AX}\Xib)$.
\end{defin}
Marginal and conditional probabilities are calculated in the standard way, e.g.~$\mc{P}(\bm{X})=\sum_{\bm{A}}\mc{P}(\bm{A}\bm{X})$ and  $\mc{P}(\bm{A}|\bm{X})=\mc{P}(\bm{A}\bm{X})/\mc{P}(\bm{X})$, and similarly for the model probabilities $P(\cdot)$. If the marginal probability distribution $\mc{P}(\Ag|\bm{X})\equiv \sum_{\Agm}\mc{P}(\bm{A}|\bm{X})$ for any compatible subset of measurement outcomes is independent of the context in which they are performed, the phenomenon is said to satisfy the condition of \emph{no-disturbance}.
\begin{defin}[No-disturbance] \label{defin:ND}
A phenomenon is said to satisfy \emph{no-disturbance} iff (i) $\mc{P}(\Ag|\bm{X})=\mc{P}(\Ag|\Xg)$ for all values of the variables $\{A_i,X_i\}$ for which those conditionals are defined, for all $\gamma \subseteq \mc{I}$ and for all $i\in \gamma$; (ii) $\mc{P}(A_i|X_i=m) = \mc{P}(A_j|X_j=m)\,\forall i,j$ for which these conditionals are defined.
\end{defin}

The second of these conditions represents the requirement that when a measurement $m$ is associated with more than one index (as can occur in KS scenarios), its marginals are independent of the index (and thus of the context).

To clarify our notation, in a scenario with three pairs of variables, the no-disturbance conditions include three constraints of the form $\mc{P}(A_1|X_1X_2X_3)=\mc{P}(A_1|X_1)$ and three of the form $\mc{P}(A_1A_2|X_1X_2X_3)=\mc{P}(A_1A_2|X_1X_2)$. In the language of causal models, these no-disturbance conditions are denoted by $(\Ag \ind \Xgm \mid \Xg)$. The decomposition axiom (\ref{decomposition}) can then be used to derive less general no-disturbance conditions for subsets of $\Xgm$. 

In Bell scenarios, when each measurement in $\bm{X}$ is space-like separated from all others, the no-disturbance condition is called the \emph{no-signalling} condition. Note that in Bell scenarios no-signalling could be defined as $\mc{P}(\Ag|\Xgm)=\mc{P}(\Ag)$. However, although this assumption is implied by our definition when the measurement settings are chosen independently, it can be violated if the choices are correlated, as is the case in general KS scenarios.

It is important to note that no-disturbance and no-signalling are defined as properties of phenomena, that is, they are defined operationally. The following definitions instead deal with properties of causal models for a phenomenon---that is, they can be understood as \emph{ontological} properties. We say that a phenomenon $\mc{P}$ \emph{satisfies} a property pertaining to causal models when there exists a causal model for $\mc{P}$ that satisfies that property, and we say that a phenomenon $\mc{P}$ \textit{violates} a property when no causal model for $\mc{P}$ satisfies that property. 

Kochen-Specker noncontextuality (KSNC)~\cite{Kochen1967} is the assumption that all measurements have a predetermined value independently of the context in which they are performed. Bell-locality, on the other hand, can be derived from two sets of assumptions\footnote{These need to be taken together with the assumption of Freedom of Choice: that measurement settings can be chosen via variables that are a priori uncorrelated with any other variables relevant to the experiment.}~\cite{Wiseman2017}: Locality and Predetermination (Bell's 1964 theorem) or Local Causality (Bell's 1976 theorem). Local Causality is a stronger notion than Locality, but weaker than the conjunction of Locality and Predetermination.

While Bell's 1964 theorem (and the KS theorem) can be resolved with the simple rejection of Predetermination, Bell's 1976 theorem requires a more radical revision on classical notions of causality. The notion of causality built into Local Causality amounts to the classical causal framework reviewed here, with the causal graph implied by relativistic causal structure.

Despite this conceptual difference, the mathematical constraints imposed by the assumptions of Bell-locality and KS-noncontextuality are essentially equivalent, and can be translated to the language of causal models through the condition of \emph{factorisability}:
\begin{defin}[Factorisability] \label{defin:FACT}
A causal model for a phenomenon $\mc{P}$ is said to satisfy \emph{factorisability} iff $\exists \Le \subseteq \Xib$ s.t.~$\forall \,\bm{A},\bm{X},$ $\mc{P}(\bm{A}|\bm{X})=\sum_{\Le} P(\Le) \prod_i P(A_i|\Le X_i)$. For KS scenarios, we require $P(A_i|\Le X_i=m) = P(A_j|\Le X_j=m)\,\forall i,j$ for which these conditionals are defined.
\end{defin}

The requirement for KS scenarios means that a measurement $m$ has the same marginal statistics in the model independently of which random variable it is associated with.
 
 By the Fine-Abramsky-Brandenburger (FAB) theorem~\cite{Fine1982, Abramsky2011}, the assumption of Kochen-Specker noncontextuality is equivalent to the existence of a factorisable model for a phenomenon satisfying no-disturbance. Bell-locality is the special case of KS-noncontextuality in a Bell scenario.
\begin{prop}[FAB theorem] \label{defin:KS-NC}
A 
phenomenon satisfies \emph{KS-noncontextuality} iff it is factorisable, i.e.~iff it has a factorisable model.
\end{prop}

The set of KS-noncontextual phenomena for each scenario is bounded by the \emph{KS inequalities}~\cite{Klyachko2008,Cabello2009}, which can be derived as the facets of a convex polytope \cite{Abramsky2012, Pitowsky1989} induced by the factorisability condition. These inequalities reduce to \emph{Bell inequalities}~\cite{Peres1999} in Bell scenarios.

\begin{figure}
    \centering
    \includegraphics[width=1\linewidth]{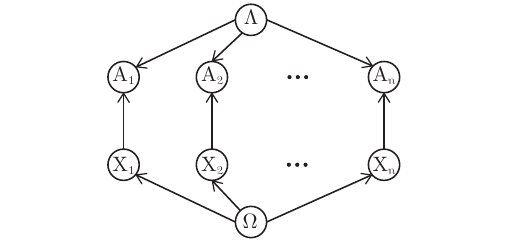}
    \caption{A canonical causal graph compatible with no-disturbance and no fine-tuning.}
    \label{fig:fig1}
\end{figure}

Every factorisable phenomenon can be modelled via a canonical causal model\footnote{Note that the term ‘canonical causal model’ is used here in a more restrictive sense than the definition given in Ref.~\cite{Jones2019}.} with a graph as given in Fig.~\ref{fig:fig1}, containing a latent variable $\Lambda$ acting as a common cause between all outcomes $A_i$ and a latent variable $\Omega$ as a common cause to all measurement choices $X_i$. In a KS scenario, $\Omega$ can be thought of as encoding the choice of context; in Bell scenarios, it is usually assumed that the choices of measurement are mutually independent, which is encoded in the typical Bell-scenario graph by ommitting $\Omega$. However, this is unnecessary: any phenomenon compatible with the graph in Fig.~\ref{fig:fig1} satisfies factorisability, as can be readily checked by applying the Causal Markov Condition to this graph and summing over $\Omega$. What is required for the derivation of Bell inequalities is that $\Omega$ is not otherwise causally connected to the other variables in the graph.

In Bell scenarios, Fig.~\ref{fig:fig1} (and thereby factorisability) is motivated by relativistic causal structure, when the different parties' events are space-like separated, plus an assumption of ``freedom of choice'' or ``statistical independence'', which can be interpreted at the causal level as the requirement that $\Lambda$ and $\Omega$ are not causally connected. Bell-locality can then be derived from an application of the Causal Markov Condition to the causal graph implied by relativistic causal structure (for a review, see \cite{Wiseman2017}). 

On the other hand, the justification of factorisability for KS scenarios, where measurements are not space-like separated, rests on more controversial grounds. It is typically derived with an assumption of outcome determinism that is arguably unjustified when formulated within the language of ontological models~\cite{Spekkens2005, Spekkens2014}. Here we show that, fortunately, this controversy can be avoided within the framework of causal models, as the condition of factorisability is implied by the principle of \emph{no fine-tuning}, or \emph{faithfulness}, a fundamental principle of causal models, without the need to invoke outcome determinism.

\begin{defin}[Faithfulness (no fine-tuning)] \label{defin:faithfulness}
A causal model $\Gamma$ is said to satisfy \emph{no fine-tuning} or be \emph{faithful} relative to a phenomenon $\mc{P}$ iff
\begin{enumerate}
    \item every conditional independence $(C\ind D|E)$ in $\mc{P}$ corresponds to a d-separation $(C\ind D|E)_d$ in the causal graph $\mc{G}$ of $\Gamma$;
    \item operational symmetries of the phenomenon are reflected in the model, rather than holding only for fine-tuned choices of model parameters.
\end{enumerate}
\end{defin}

Consider a phenomenon that is known to satisfy the C.I.~relation $(A\ind B \mid C)$ (corresponding, for example, to a no-signalling condition). If the causal structure does not satisfy the d-separation $(A\ind B \mid C)_d$, then the observed conditional independence can only arise due to specially \emph{fine-tuned} values of the causal parameters. These fine-tuned parameters act to ``hide'' causal connections (for example, faster-than-light causation), creating the illusion of a C.I.~relation at the operational level. A faithful causal model is then best understood to be a causal model with no hidden causal connections.

The symmetry condition in Def.~\ref{defin:faithfulness}.2 is an extension to Pearl's notion of faithfulness, but is essential for contextuality scenarios, as we will see in the final step of the proof of Theorem 1. The specific implication of this assumption that we need is that if the marginals of a phenonemon are symmetric with respect to exchange of labels associated with a measurement $m$ (e.g.~if $\mc{P}(A_i|X_i=m) = \mc{P}(A_j|X_j=m)$ for some $i,j$) the model should satisfy the same symmetry. This assumption is analogous to the notion of ``operational no fine-tuning'' recently introduced by Catani and Leifer~\cite{Catani2020}. Its requirement for KS scenarios highlights a subtle but important distinction in the assumption of no-fine-tuning required for KS scenarios vs Bell scenarios.

The motivation for no fine-tuning is analogous to that for {\it Leibniz's principle of the identity of indiscernibles}~\cite{Spekkens2019}, which states that a theory should avoid postulating distinctions at the ontological level that are not reflected in operational distinctions. It can also be understood as the methodological principle underlying Einstein's principles of relativity and of equivalence~\cite{Spekkens2019}.

In light of the above discussion, the violation of a Bell-KS inequality in a phenomenon $\mc{P}$ implies that (i) either the causal graph underlying the phenomenon does not have the form of the canonical causal graph in Fig.~\ref{fig:fig1}, or (ii) the classical causal model formalism needs to be rejected or modified so that this graph does not imply factorisability (e.g.~ as in the program of quantum causal models~\cite{Cavalcanti2014, Henson2014, Pienaar2015, Chaves2015, Costa2016, Allen2016, Barrett2019}). For Bell scenarios, the causal graph is motivated by relativity, and proposed resolutions of the type (i) above include violations of relativistic causality (e.g.~in Bohmian mechanics~\cite{Bohm1}), retrocausality~\cite{Price2008} or superdeterminism~\cite{tHooft2007}. A priori, these alternative causal structures seem objectionable for different reasons, but they all share a common property: they require fine-tuning within a classical causal model. A natural question is therefore whether another modified causal structure, however exotic, could reproduce the violation of Bell inequalities while avoiding this objection. For KS scenarios, on the other hand, the canonical causal graph in Fig.~\ref{fig:fig1} is not directly motivated by relativity, and the question is how can factorisability be motivated for these scenarios at all. The present result completes the partial results of \cite{Wood2015,Cavalcanti2018} and resolves both of these questions at once, for arbitrary Bell-KS scenarios: it establishes that any classical causal model for such scenarios that is faithful to the no-disturbance conditions implies factorisability.

\section{Main result}\label{sec:main}

In \cite{Cavalcanti2018} (following \cite{Wood2015}), it was shown that no-fine-tuning leads to KS-noncontextuality for any phenomenon satisfying no-disturbance. The proof however was restricted to contextuality scenarios with two measurements per context (and bipartite Bell scenarios as a special case). Here we show that this result holds in general scenarios with arbitrary numbers of parties or measurements per context.

\begin{thm} \label{thm:main} Every phenomenon satisfying no-disturbance in an arbitrary contextuality scenario that has a faithful causal model is factorisable.
\end{thm}

Theorem~\ref{thm:main} leads to the following immediate corollaries:
\begin{cor} \label{cor:1} No fine-tuning and no-disturbance (no-signalling) imply KS noncontextuality (Bell-locality) in arbitrary scenarios.
\end{cor}
\begin{cor} \label{cor:2} Every classical causal model that reproduces the violation of a Bell-KS inequality in a no-disturbance phenomenon in an arbitrary Bell-KS scenario requires fine-tuning.
\end{cor}

\section{Example scenario based on the Peres-Mermin square}\label{sec:example}
\begin{figure}[H]
    \centering
    \includegraphics[width=1\linewidth]{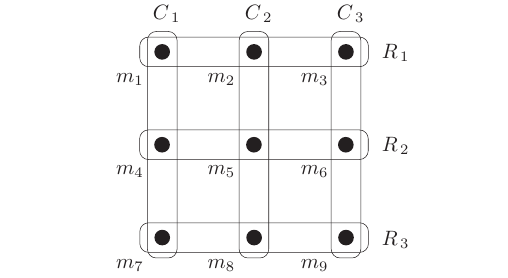}
    \caption{Compatibility hypergraph for the nine binary outcome measurements in the Peres-Mermin square. $R_1$, $R_2$ and $R_3$ represent the set of measurements in rows 1, 2 and 3 respectively. Likewise, $C_1$, $C_2$ and $C_3$ represent the set of measurements in columns 1, 2 and 3.}
    \label{fig:fig2}
\end{figure}

In~\cite{Cavalcanti2018}, it was shown how the causal framework for contextuality can be mapped onto a three-measurement scenario with two measurements per context introduced by Liang, Wiseman, and Spekkens \cite{LWS}. Here, we demonstrate an example with three measurements per context based upon the the Peres-Mermin square \cite{Peres1991,Mermin1993}. This scenario contains nine measurements $\mc{M}=\{m_1,m_2,\dots,m_9\}$ with binary outcomes $\mc{O}=\{-1,1\}$. The compatibility structure can be conveniently represented by the hypergraph in Fig.~\ref{fig:fig2}~\cite{Krishna2017}.

Rows and columns represent measurement contexts---that is, they are jointly measurable. Formally, the compatibility structure is denoted by~$\mc{C}=\{R_1,R_2,R_3,C_1,C_2,C_3\}$. For this scenario, $n=3$, as each measurement context contains at most three measurements. In each run of the experiment, $\bm{X}=\{X_1,X_2,X_3\}$ can take any triplet of values from $\mc{C}$, and $A_{i}\in\bm{A}=\{A_1,A_2,A_3\}$ can take values from $\mc{O}=\{-1,1\}$.
Consider a phenomenon $\mc{P}=(\bm{AX})$ that satisfies the no-disturbance relations $\mc{P}(\Ag|\bm{X})=\mc{P}(\Ag|\Xg)$. In this scenario, $\gamma=\{1,2,3,
\{1,2\},
\{1,3\},
\{2,3\}
\}$, and so there is a no-disturbance relation for each member of $\gamma$. From Theorem 1, any faithful classical causal model for this phenomenon must satisfy KS-noncontextuality. In this scenario, KS-noncontextuality implies the inequality \cite{Cabello2008}:
\begin{multline}\label{eqn:ineq}
    \expect{\text{KS}} = \expect{R_1} + \expect{R_2} + \expect{R_3} \\
    + \expect{C_1} + \expect{C_2} - \expect{C_3} \leq 4.
\end{multline}

Quantum theory predicts a state-independent violation of this inequality by two qubits, where the measurement scenario is represented by the following array of Pauli spin matrices,
\begin{equation}
    \begin{array}{ccc}
    \sigma_x^{(1)} & \sigma_x^{(2)} & \sigma_x^{(1)} \otimes \sigma_x^{(2)} \\
    \sigma_y^{(2)} & \sigma_y^{(1)} & \sigma_y^{(1)} \otimes \sigma_y^{(2)} \\
    \sigma_x^{(1)} \otimes \sigma_y^{(2)} & \sigma_x^{(2)} \otimes \sigma_y^{(1)} & \sigma_z^{(1)} \otimes \sigma_z^{(2)}.
\end{array} 
\end{equation}

Quantum theory predicts, for any state, $\expect{\text{KS}}=6$, thus violating inequality~(\ref{eqn:ineq}). Every classical causal model for a no-disturbance phenomenon which violates inequality~(\ref{eqn:ineq}) therefore requires fine-tuning.

\section{Proof of Theorem 1}\label{sec:proofs}
\begin{figure}[H]
	\centering
	\includegraphics[width=0.9\linewidth]{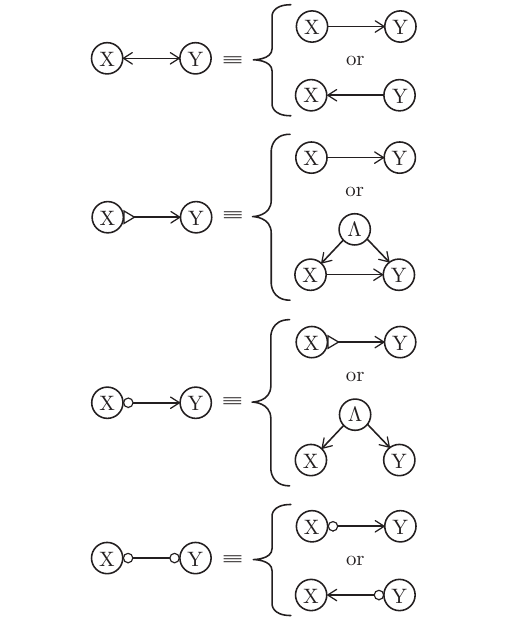}
	\caption{Shortcut graphical notations for causal connections between X and Y.}
	\label{fig:fig3}
\end{figure}

To aid in the proof of Theorem 1, we introduce the graphical notations in Fig.~\ref{fig:fig3} to represent sets of causal connections. A diagram using these shortcut notations represents the set of all DAGs compatible with all shortcut notations. A dashed line represents a connection of the type indicated or no connection. 
The proof will make use of the following Lemma:
\begin{lem}\label{lem:chain}
Let a \emph{chained graph} be a graph of the form below (Fig. \ref{fig:fig4}), where \textbf{A,B,C,D} represent sets of vertices, and the connections between two such sets represent possible connections between any pairs of elements in each set. Note that all paths between elements of non-adjacent sets go through the intermediate set, e.g. all paths from A to C go through at least one element of B.
\begin{figure}[H]
	\centering
	\includegraphics[width=1\linewidth]{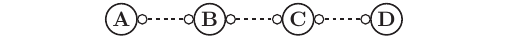}
	\caption{A chained graph $\mc{V}_c$.}
	\label{fig:fig4}
\end{figure}
If $\mc{V}_c$ satisfies $(\bm{A}\ind \bm{C}\mid \bm{B})_d$, then it also satisfies $(\bm{A}\ind \bm{CD}\mid \bm{B})_d$ and $(\bm{A}\ind \bm{D}\mid \bm{BC})_d$.
\end{lem}

\paragraph{Proof of Lemma~\ref{lem:chain}.}
To satisfy $(\bm{A}\ind \bm{C}\mid \bm{B})_d$, $\bm{B}$ must block all paths between $\bm{A}$ and $\bm{C}$, from which it follows that $\bm{B}$ blocks all paths between $\bm{A}$ and $\bm{D}$ (as all paths must pass through at least one element of $\bm{B}$, and through an element of $\bm{C}$ before reaching $\bm{D}$). Likewise $\bm{B}$ blocks all paths between $\bm{A}$ and $\bm{C}\bm{D}$. We can then write $(\bm{A}\ind \bm{D}\mid \bm{B})_d$ and $(\bm{A}\ind \bm{CD}\mid \bm{B})_d$. From the second of these and the weak union axiom \eqref{weakunion} we can then derive $(\bm{A}\ind \bm{D}\mid \bm{BC})_d$.\QED

The no-disturbance condition, when combined with the assumption of no fine-tuning, leads to the d-separation conditions
\begin{equation}\label{nodd}
(\Ag\ind \Xgm \mid \Xg)_d.
\end{equation}
The rest of the proof proceeds by deriving a set of d-separation conditions that must be obeyed by any causal model satisfying~(\ref{nodd}). These d-separation conditions then imply new C.I.~relations in any joint distribution compatible with any faithful causal graph satisfying no-disturbance, which will be shown to imply factorisability in the joint distribution for any number of parties or measurements per context.

\paragraph{Step 1a.}
The class of DAGs we need to consider are those that include latent variables as common causes for observable variables or direct causal connections between them. There is no need to consider latent variables as intermediaries or common effects between variables, since those have no effect on the allowed probability distributions over the observable variables.

\begin{figure}
	\centering
	\includegraphics[width=1\linewidth]{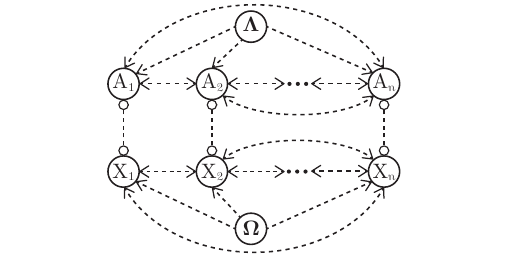}
	\caption{Remaining class of DAGs after Step 1a.}
	\label{fig:fig5}
\end{figure}

From (\ref{nodd}), we can exclude any direct causal link or common cause between $\{A_i,\Xe\}$ for all $ i\in \mc{I}=\{1,2,...,n\}$. In particular, this excludes the possibility of a common cause between any two or more outcomes and a setting, or between any two or more setting and an outcome. This leaves us with the possibility of any causal link between $\{A_i,X_i\}$, $\{A_i,\Ae\}$ and $\{X_i,\Xe\}$ as shown in Fig.~\ref{fig:fig5}. Without loss of generality, we then introduce $\Le$ and $\Om$ as the sets of latent variables potentially acting as common causes between all outcomes and all settings, respectively, and the above remark implies that $\Le$ and $\Om$ are not directly causally connected. Considering intermediate latent variables as joint intermediaries would not allow for more general phenomena, and would not create any causal paths between the observable variables that are not already included in Fig.~\ref{fig:fig5}.

\begin{figure}
    \centering
    \includegraphics[width=1\linewidth]{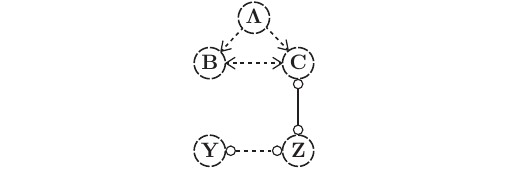}
	\caption{Shortcut representation of the class of DAGs in Fig.~\ref{fig:fig5}. Dashed circles represent the possibility of an empty set. $\bm{B}\subseteq \bm{A}$: all members of $\bm{A}$ with no causal connection to $\bm{X}$; $\bm{C}\subseteq \bm{A}$: all members of $\bm{A}$ with some causal connection to $\bm{X}$. A connection between two sets represents all possible connections of the type indicated between members of each node.}
	\label{fig:fig6}
\end{figure}

\paragraph{Step 1b.}
For every DAG represented by Fig.~\ref{fig:fig5}, and without loss of generality, all members of $\bm{A}$ and $\bm{X}$ can be grouped into subsets depending on the existence of certain causal connections, as shown in Fig~\ref{fig:fig6}. All members of $\bm{A}$ with no direct causal connection to any member of $\bm{X}$ are denoted by the subset $\bm{B}$. Remaining members of $\bm{A}$ are denoted by the subset $\bm{C}$. Likewise, all members of $\bm{X}$ with no direct causal connection to any member in $\bm{A}$ are denoted by $\bm{Y}$, while the remaining members are denoted by $\bm{Z}$.

\begin{figure}
    \centering
    \includegraphics[width=1\linewidth]{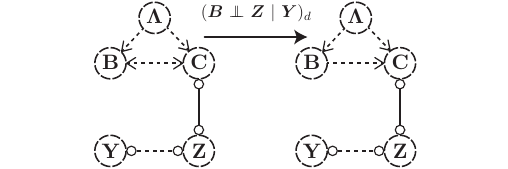}
	\caption{Elimination of direct links from $\bm{C}$ to $\bm{B}$.}
	\label{fig:fig7}
\end{figure}

\paragraph{Step 2a.}From (\ref{nodd}), we can derive $(\bm{B}\ind \bm{Z}\mid\bm{Y})_d$. Note that any path between $\bm{B}$ and $\bm{Z}$ must pass through at least one element of $\bm{C}$. Therefore, for any such path, $\bm{C}$ acts as a middle node that is not in $\bm{Y}$. For $\bm{B}$ to be d-separated from $\bm{Z}$ given $\bm{Y}$, $\bm{C}$ must act as a collider in any path between $\bm{B}$ and $\bm{Z}$. Since every member of $\bm{C}$ has a connection to one and only one member of $\bm{Z}$, any member of $\bm{C}$ with a direct causal connection to $\bm{B}$ would be a non-collider middle node between $\bm{B}$ and $\bm{Z}$.  Thus, direct connections from $\bm{C}$ to $\bm{B}$ would violate $(\bm{B}\ind \bm{Z}\mid\bm{Y})_d$, and are excluded, as shown in Fig.~\ref{fig:fig7}.
\paragraph{Step 2b.}From Fig.~\ref{fig:fig7} we see that $\bm{Y}$ cannot act as a middle node in paths between $\bm{B}$ and $\bm{Z}$. Thus $(\bm{B}\ind \bm{Z}\mid\bm{Y})_d$ implies $(\bm{B}\ind \bm{Z})_d$. Therefore $\bm{B}$ is d-separated from $\bm{Z}$ given any variable that is not a collider in a path between them. As $\Le$ satisfies this condition, we find that
\begin{equation}\label{d1}
(\bm{B}\ind \bm{Z}\mid\Le)_d.
\end{equation}

Using (\ref{nodd}) again, we can write the d-separation condition $(\bm{C}\ind \bm{Y}\mid\bm{Z})_d$. From the symmetry axiom (\ref{symmetry}), this can be rewritten as $(\bm{Y}\ind \bm{C}\mid\bm{Z})_d$. From Lemma \ref{lem:chain}, it follows that all graphs in Fig.~\ref{fig:fig7} compatible with $(\bm{Y}\ind \bm{C}\mid\bm{Z})_d$ must also satisfy $(\bm{Y}\ind \bm{CB}\Le\mid\bm{Z})_d$. From the weak union axiom (\ref{weakunion}), this can be rewritten as $(\bm{Y}\ind \bm{CB}\mid\bm{Z}\Le)_d$. Reapplying the symmetry axiom (\ref{symmetry}) and rewriting $\bm{BC}=\bm{A}$, we arrive at the condition
\begin{equation}\label{d2}
    (\bm{A}\ind \bm{Y}\mid\bm{Z}\Le)_d.
\end{equation}

\begin{figure}
    \centering
    \includegraphics[width=1\linewidth]{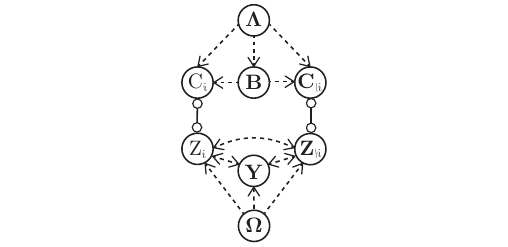}
    \caption{Representation of the set of graphs with $\bm{C}$ separated into $C_i$ and $\Ce$ and $\bm{Z}$ separated into $Z_i$ and $\Ze$.}
    \label{fig:fig8}
\end{figure}

\paragraph{Step 3a.}
Now we consider the causal connections between two arbitrary variables $\{C_i,C_j\}\in \bm{C}$. From (\ref{nodd}) and the decomposition axiom (\ref{decomposition}), we can write the condition $(C_j \ind Z_i\mid Z_j)_d$. Consider the path $(Z_i - C_i- C_j)$ for arbitrary $i,j$. For this path to be blocked by $Z_j$, the middle node $C_i$ must be a collider. Thus, we can eliminate direct connections $C_i\rightarrow C_j$ between any two members of $\bm{C}$, as shown in Fig. \ref{fig:fig8}.

\paragraph{Step 3b.}
We now consider what d-separation conditions can be found between members of $\bm{C}$, as this will be required for the final step. Considering Fig.~\ref{fig:fig8}, all paths between $C_i$ and $\bm{C}_{\setminus i}$ can be divided in two classes: (i) those paths that go through $Z_i$ (the bottom half of Fig.~\ref{fig:fig8}) and (ii) those that go through $\bm{B}\bm{\Lambda}$ (the top half). 
Consider the paths in (i). From (\ref{nodd}) and the decomposition axiom (\ref{decomposition}), we can write $(C_i\ind \Ze \bm{Y} \mid Z_i)_d$. We can see from Fig.~\ref{fig:fig8} that every path between $C_i$ and $\Ze \bm{Y}$ that contains a collider in $Z_i$ violates this condition. Thus any such path must have a chain or fork with $Z_i$ as the middle node. Thus $Z_i$ blocks, as a chain or fork, all paths between $C_i$ and $\Ce$ that go through $Z_i$. Therefore $\bm{Z}$ blocks all paths in (i) between $C_i$ and $\Ce$.

The paths in (ii) are blocked by conditioning on $\bm{B}\Le$, as all such paths are chains or forks with $\bm{B}$ and/or $\Le$ as a middle node. Thus, we find that all DAGs in Fig.~\ref{fig:fig8} that satisfy condition (\ref{nodd}) also satisfy
\begin{equation}\label{d3}
    (C_i\ind \Ce\mid \bm{ZB\Lambda})_d.
\end{equation}

Any path through $\Le$ is blocked by $\Le$ because it is a fork. From $(C_i\ind \Ze\mid Z_i)_d$ we can then write 
\begin{equation}\label{d4}
    (C_i\ind \Ze\mid Z_i\Le)_d,
\end{equation}
as conditioning on $\Le$ cannot make $C_i$ and $\Ze$ dependent.

\begin{figure}[H]
    \centering
    \includegraphics[width=1\linewidth]{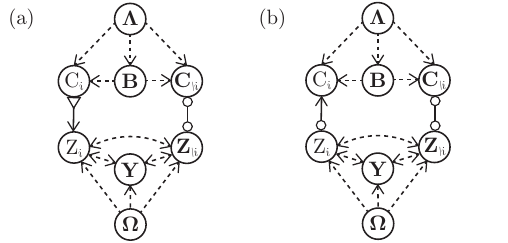}
    \caption{Two sub-graphs of Fig.~\ref{fig:fig8}, where (a) considers a direct connection from $C_i$ to $Z_i$ with or without a common cause, and (b) excludes a connection from $C_i$ to $Z_i$. Together these graphs account for all graphs in Fig.~\ref{fig:fig8}.}
    \label{fig:fig9}
\end{figure}

\paragraph{Step 3c.}
This step consists of deriving the d-separation condition $(\Le\ind\bm{X})_d$. It is not necessary to consider paths through $\bm{B}$, as every path from $\Le$ to a member of $\bm{Z}$ must pass through a member of $\bm{C}$. For every member $C_i$ of $\bm{C}$, we can separate the graphs in Fig.~\ref{fig:fig8} into the two sub-graphs shown in Fig.~\ref{fig:fig9}.

Now consider the class of graphs in~Fig.~\ref{fig:fig9}a. The conditions $(\bm{B}\ind \bm{Z} \mid \bm{Y})_d$ and $(\Ce\ind Z_i \mid \Ze)_d$ exclude the possibility of a common cause between $C_i$ and any member of $\bm{B}$ or $\Ce$, as well as direct connections from $\bm{B}$ to $C_i$, as shown below in Fig.~\ref{fig:fig10}. Therefore there are no paths of the type $\Le-C_i-Z_i$ when there is a direct connection from $C_i$ to $Z_i$.
\begin{figure}[H]
    \centering
    \includegraphics[width=1\linewidth]{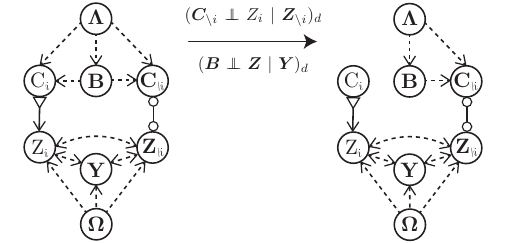}
    \caption{Elimination of causal connections from Fig.~\ref{fig:fig9}a.}
    \label{fig:fig10}
\end{figure}

The path $\Le-C_i-Z_i$ in~Fig.~\ref{fig:fig9}b is blocked by the empty set, as $C_i$ acts as a collider with no descendants. Since every path between $\Le$ and $\bm{Z}$ includes a sub-path of the form of $\Le-C_i-Z_i$ in Fig.~\ref{fig:fig9}a or Fig.~\ref{fig:fig9}b, we can then write
\begin{equation}\label{d5}
    (\Le\ind \bm{X})_d,
\end{equation}

\paragraph{Step 4.}
The d-separation conditions derived in (\ref{d1}), (\ref{d2}), (\ref{d3}), (\ref{d4}) and (\ref{d5}) imply the corresponding C.I.~conditions
\begin{align}
(\Le &\ind \bm{X}) \label{e0}\,,\\
(\bm{B}&\ind \bm{Z}\mid\Le) \label{e1}\,,\\
(\bm{A}&\ind \bm{Y}\mid\bm{Z\Le}) \label{e2}\,,\\
(C_i&\ind \Ce\mid\bm{ZB\Le})\label{e3}\,,\\
(C_i&\ind \Ze\mid Z_i\Le) \label{e4}\,.
\end{align}
From the definition of conditional probability, we can write the observable joint distribution as
\begin{equation}
\mc{P}(\bm{A}\mid\bm{X})=\sum_{\Le\, \bm{\Omega}}P(\bm{A \bm{\Omega}}\mid\bm{X}\Le)P(\Le\mid\bm{X})\,.
\end{equation}
Summing over $\bm{\Omega}$, writing $\bm{X}=\bm{YZ}$ and using (\ref{e0}), we can write
\begin{equation}
\mc{P}(\bm{A}\mid\bm{X})=\sum_{\Le}P(\bm{A}\mid\bm{YZ}\Le)P(\Le)\,.
\end{equation}
Substituting $\bm{A}=\bm{BC}$, and using (\ref{e1}) and (\ref{e2}),
\begin{equation}
\mc{P}(\bm{A}\mid\bm{X})=\sum_{\Le}P(\bm{C}\mid\bm{ZB}\Le)P(\bm{B}\mid\Le)P(\Le)\,.
\end{equation}
Now note from~Fig.~\ref{fig:fig8} that no observable variable outside $\bm{B}$ can have a direct causal link to $\bm{B}$. This is reflected in the equation above by the fact that $\bm{B}$ is only dependent on $\Le$. Without loss of generality, we can thus let $\Le$ determine $\bm{B}$, as any phenomenon that is compatible with a model of this form is also compatible with a model where $\Le$ determines $\bm{B}$ (after suitable fine-graining). Any information about $\bm{B}$ is then known given $\Le$. Since $\Le$ determines $\bm{B}$, $P(\bm{C}\mid\bm{ZB}\Le) = P(\bm{C}\mid\bm{Z}\Le)$. Using the definition of conditional probability,
\begin{equation}
P(\bm{C}\mid\bm{Z}\Le)=\prod_{j}P(C_j\mid\bm{C}\setminus \{C_1,C_2,\dots,C_{j}\}\bm{Z}\Le).
\end{equation}
From (\ref{e3}), all $C_j\in\bm{C}$ are independent given $\bm{Z}\Le$. From (\ref{e4}), all $C_j\in\bm{C}$ are independent of $\Zj$ given $Z_j\Le$. Thus
\begin{equation}
P(\bm{C}\mid\bm{Z}\Le)=\prod_{j}P(C_j\mid Z_j\Le)\,.
\end{equation}
Applying this procedure to $P(\bm{B}\mid\Le)$, and since by definition $\Le$ determines $\bm{B}$, we can similarly write
\begin{equation} \label{eq:Bk}
P(\bm{B}\mid\Le)=\prod_{k}P(B_k\mid\Le) \,.
\end{equation}
We can finally write the observable joint distribution for any phenomena satisfying no-disturbance and no-fine-tuning as:
\begin{align}
\mc{P}(\bm{A}\mid\bm{X})=\sum_{\Le}P(\Le)\prod_{j,k} P(C_j\mid Z_j\Le) P(B_k\mid\Le) \,.
\end{align}
For Bell scenarios, where each measurement is performed by a single party, this is a factorisable model, completing the proof.

\paragraph{Step 5.}
For KS scenarios, a further step is needed, to justify that $P(A_i|\Le X_i=m) = P(A_j|\Le X_j=m)$, as noted below Prop.~\ref{defin:KS-NC}. This condition follows from Def.~\ref{defin:faithfulness}.2 by noting that (from Def.~\ref{defin:ND}) the marginals of a no-disturbance phenomenon are symmetric with respect to exchange of labels associated with a measurement $m$ (i.e. $\mc{P}(A_i|X_i=m) = \mc{P}(A_j|X_j=m)$ for all $i,j$ in which $m$ appears). This completes the proof for KS scenarios, and highlights a key distinction between KS and Bell scenarios.

\QED

\section{Conclusion}\label{sec:conclusion}
In summary, we have shown that Kochen-Specker contextuality and Bell-nonlocality, in fully general scenarios with arbitrary numbers of measurements per context or parties, and arbitrary numbers of outcomes per measurement, can both be understood as phenomena for which it is impossible to construct a faithful classical causal model. This means that these key quantum phenomena can be understood in a unified way as violations of the classical framework of causality.

This result has several important consequences. Firstly, from a foundational perspective, it generalises the results of \cite{Wood2015, Cavalcanti2018}, confirming that this relationship between fine-tuning and Bell-KS inequality violations is fully general, and not an artefact of the simplest scenarios. This adds extra motivation for the program of quantum causal models~\cite{Cavalcanti2014, Henson2014, Pienaar2015, Chaves2015, Costa2016, Allen2016, Barrett2019}, in which the classical framework of causality is extended into a framework where quantum correlations can be potentially explained \emph{without} fine-tuning, thus removing the objectionable property that, according to the present result, is required of classical causal models for all Bell-KS correlations.

Secondly, as alluded to in the introduction, and in \cite{Cavalcanti2018}, this result gives a general motivation for the idea of quantifying quantum advantage via fine tuning, as it shows that this is a property of all classical simulations of phenomena displaying Bell-KS contextuality---key resources for quantum communication and computation protocols. It also puts in a new light previous results about simulating Bell correlations with the aid of extra communication, such as the seminal work of Toner and Bacon~\cite{Toner2003} and subsequent results.

Another avenue for further research is to extend the principle of no fine-tuning to accommodate phenomena that do not satisfy no-disturbance, as is the case in non-ideal experiments. In~\cite{Cavalcanti2018}, a \emph{generalised principle of no fine-tuning} was proposed, whereby a causal model should not allow causal connections stronger than needed to explain the observed deviations from the no-disturbance condition. A recent work~\cite{Jones2019} has implemented a version of this principle, and shown that models that satisfy this property (dubbed ``M-noncontextuality'') are equivalent to models that satisfy the property of CbD-noncontextuality defined in the ``Contextuality by Default'' approach~\cite{Dzhafarov2016}. However, the class of causal models considered in~\cite{Jones2019} is not as general as the ones considered here: it only considers a minimal relaxation from the default causal structure, allowing for causal influences from the contexts to the measurement outcomes. This excludes by fiat a large class of candidate causal models. It would be interesting to know whether this relationship between generalised no-fine-tuning and CbD-contextuality holds in general, and whether this can lead to robust experimental tests.

Finally, we note that although this result unifies Bell-nonlocality and Kochen-Specker contextuality as violation of no-fine-tuning in classical causal models, this does not imply that these phenomena represent all forms of quantum violations of classical causality. It has recently been shown~\cite{Renou2019}, for example, that quantum correlations can violate the classical constraints on a ``triangle scenario''~\cite{Fritz2012} even in the absence of any choice of measurement settings, a phenomenon dubbed ``quantum nonlocality without inputs''. Since there are no choices of settings, there are no C.I. relations associated with no-disturbance conditions in those scenarios. Therefore the quantum violations of those kinds of scenarios are not instances of Bell-nonlocality nor KS-contextuality, but some fundamentally different kind of nonclassicality. Understanding the nature of this distinction is an interesting question for further study.

\begin{acknowledgments}
Primary acknowledgements go to Matt Jones for extensive and thoughtful feedback on an earlier version of the manuscript, leading to important changes to the final definitions and results. We also acknowledge Shane Mansfield and two anonymous referees for useful and constructive feedback. This work was supported by the Australian Research Council Future Fellowship FT180100317, and grant number FQXi-RFP-1807 from the Foundational Questions Institute and Fetzer Franklin Fund, a donor advised fund of Silicon Valley Community Foundation.
\end{acknowledgments}

\bibliography{fine-tuning.bbl}

\end{document}